# MIXED-EFFECTS METHODS FOR SEARCH AND MATCHING RESEARCH

John M. Abowd and Kevin L. McKinney[1]

August 28, 2023

Keywords: mixed-effects models, generalized linear models

JEL Classifications: J3, C5

Abstract

We study mixed-effects methods for estimating equations containing person and firm effects. In economics such models are usually estimated using fixed-effects methods. Recent enhancements to those fixed-effects methods include corrections to the bias in estimating the covariance matrix of the person and firm effects, which we also consider.

---

[1] Author information: Abowd, Cornell University, Ithaca, NY 14853 USA, john.abowd@cornell.edu; McKinney, U.S. Census Bureau, Washington, DC 20233 USA, kevin.l.mckinney@census.gov. This paper was originally prepared for the Search and Matching Conference in honor of Jean-Marc Robin held at Sciences Po, Paris, France, December 20, 2022. We thank the participants in that conference, the editor and referee for very helpful comments. The views expressed in this paper are the authors' and not those of the U.S. Census Bureau. The research was authorized under Project ID: 6000266. Statistical results were cleared for public release under clearance number: CBDRB-FY23-CED006-0005. This version: August 28, 2023. Forthcoming in *Revue Economique*.

# Introduction

The decomposition of log wages or earnings into linearly additive effects due to the employee (person), employer (firm), and control variables (*X*) was introduced to economists by Abowd, Kramarz and Margolis [1999b] and is now familiarly called the AKM model. Statisticians have used similar methods for seven decades, where the model is known as an analysis of covariance with two (or more) high-dimensional main effects whose levels must be estimated [Henderson, 1950, 1953; Searle, 1971]. Original applications in biostatistics arose in animal husbandry, for example, because milk yield and characteristics (e.g., butterfat content) were predicted by the "person" effect (dairy cow) and the "firm" effect (the cow's genetic ancestry, in particular the bull that inseminated the cow's mother). High-precision computational techniques for estimating the "person" and "firm" effects in such models have been used for decades. In the early 2000s, the computational methods used in statistics could not accommodate high dimensional regressor variables, which led to the development of custom programming for the AKM fixed-effects specification [Abowd, Creecy and Kramarz, 2002] (ACK hereafter) that used a bespoke library of sparse matrix routines to efficiently estimate the fixed-effects model originally proposed by AKM.

Econometricians have focused almost exclusively on the properties of the fixed-effects estimator used in ACK. They have noted issues with the identification of effects (statisticians call these the estimable functions), particularly the person effects, when there is limited mobility in the estimation sample, originally noted in AKM, who did correct their estimate of the person effects, and thoroughly studied in Andrews, Schank and Upward [2008]. Recently, Kline, Saggio and Sølvsten [2020a] (KSS hereafter) proposed leave-one-out estimators for non-linear functions of the fixed person and firm effects that permit unbiased estimation of these functions within the estimation sample. Such non-linear functions include the covariance matrix of the estimated person and firm effects. As implemented in their code, the KSS estimator is not an exact fixed-effects estimator; however, using methods like ACK to pre-estimate the regressors and applying standard results from partitioned regression, the KSS code delivers an exact fixed-effects estimator. When the input data set is sampled, implying that the complete employer-employee graph cannot be used, mixed effects methods may be more reliable. Bonhomme, Holzheu, Lamadon, Manresa, Mogstad and Setzler [2022] perform exercises that also incorporate some mixed-effect methods, concluding that "limited mobility bias is severe, and that bias correction is important" (p. 292). We fundamentally agree with these conclusions. Our goal is to provide practical guidance when researchers wish to use the estimated person and firm effects in other models.

Since well-implemented mixed-effects methods do not suffer from limited mobility bias, the conditional modes may not require bias correction when used in some post-processing. Our recommendation is to incorporate mixed-effects methods more broadly in the estimation of AKM models, including in domains where there are several high-dimension main effects (or interactions), many regressors, and a desire to use the estimated levels of these effects for additional statistical analysis. Mixed-effects methods are a viable dimension-reducing method that permits analyses in the same spirit as the original AKM specification. Our recommendation is tempered by the inability to directly estimate the covariance of person and firm effects in the mixed-effects specification.

Statisticians specializing in models with the same statistical structure as AKM have now produced very efficient computational libraries with a wide variety of estimation options including fixed-effects estimation—the framework originally proposed in AKM and fully implemented in ACK, random-effects estimation—the framework in which all parameters are treated as random, and mixed-effects

estimation—the framework in which some parameters are treated as fixed and others as random [Bates, Mächler, Bolker and Walker, 2015; Bates, 2023; Bates and Calderón, 2023]. This computational framework considers the estimable linear functions of the unconstrained person and firm effects, known in econometrics as the identifiable parameters, as the estimands of interest. Abowd and Kramarz [1999c] laid out the connection between the fixed-effects formulation in AKM and the mixed-effects formulation used in statistics. They showed that the statistical assumptions regarding the design matrices for all effects and the residual are identical in the two formulations. Woodcock [2008] used the mixed-effects specification with the addition of a random match effect.

In fixed-effects methods, the estimable functions depend upon the network connectivity of the bipartite graph of the persons and firms, usually represented by an adjacency matrix. For tractability, most analysts now use the largest connected subcomponent of this graph and form the estimable functions by constraining the person and firm effects to sum to a known constant (usually zero). In random and mixed-effects estimation, the estimable functions are determined by the properties of the design matrices and the statistical assumptions on the random effects. Once those are specified, the properties of the person-firm graph no longer constrain the estimation. That is, the estimated person and firm effects can be computed for all entities in the graph regardless of its connectivity. The only requirement is that every person work for at least one employer and every employer have at least one employee. Of course, the statistical properties of the estimated person and firm effects do depend on connectivity, but the estimable functions/identifiable parameters do not (see the Appendix, Section B). On the other hand, random and mixed-effects methods require careful attention to the preparation of the estimation sample. In general, the statistical parameters (regression coefficients and the covariance matrix of the random effects) must be estimated from a sample of the bipartite graph. The estimated person and firm effects, known as estimated "best" linear unbiased predictors, but more correctly labeled the conditional modes of the random effects, can be computed for every person and firm in the entire graph, not just those included in the estimation sample.[2]

Economists have published many articles that take up the question of how to interpret the result that there is a statistical effect on log earnings directly attributable to the firm. This question is thoroughly studied in Card, Heining and Kline [2013] (CHK hereafter), who proposed an event-study framework for establishing that the firm effect operates statistically independent of the person effect. When coupled with the KSS estimator, the CHK framework provides a large set of econometric tools for applying the fixed-effects AKM specification to study labor market heterogeneity.

---

[2] This footnote defines some of the terms used in statistics that may be unfamiliar to economists. The design matrix consists of the regressors and the matrices of indicator variables that associate a person and firm effect to a particular observation. In fixed-effects estimation, the estimated person and firm effects are computed using any convenient algorithm to solve the normal equations without computing an inverse. There are an infinite number of solutions. Identification is achieved by imposing one restriction each on these estimated person and firm effects—usually, that they sum to zero across all jobs. The mixed-effects counterpart to the normal equations in fixed-effects estimation is the Henderson equations. In the standard implementation [e.g., Bates et al., 2015], all random effects are assumed normal to form a likelihood function. The log likelihood function, multiplied by -2, is called the deviance function. Conditional on the mixed-effects variance parameters, the mode of this likelihood function is the solution to the Henderson equations, which is computed using a convenient algorithm that, as in fixed-effects estimation, does not require matrix inversion. When the estimated conditional modes are substituted into the deviance function, the resulting function is called the profiled deviance. The profiled deviance depends only on the variance parameters and is maximized by numerical search. Mixed-effects estimation alternates between solving the Henderson equations for the conditional modes (given the variance parameters) and maximizing the profiled deviance until convergence. The reported conditional modes are those attained at the values of the variance parameters that maximize the profiled deviance.

The economics literature has many threads using AKM specifications. The one most salient for this conference honoring Jean-Marc Robin are papers that take the AKM statistical framework and develop structural search and matching models for interpreting the person and firm effects [Postel-Vinay and Robin, 2002; Mortensen, 2005; Cahuc, Postel-Vinay and Robin, 2006]. These economic models have also been expressed in dimension-reducing forms that permit direct assessment of their implications for the structure of the latent effects presumed to drive the economic equilibrium. See especially Abowd, McKinney and Schmutte [2019] and Bonhomme, Lamadon and Manresa [2022]. This paper does not explicitly consider dimension-reducing methods because we wish to focus on methods where the estimated person and firm effects can be associated with the original job data and used for post-processed statistical analysis. We note that mixed-effect estimation is inherently dimension reducing because the set of parameters subject to estimation error includes only the fixed-effects coefficients on regressors and the covariance matrix of the random effects. Because the regression coefficients and covariance matrix can be estimated with high precision, they do not contribute appreciably to the error in the conditional modes. The interpretation of the error in estimating conditional modes depends primarily on the superpopulation model assumed to generate the bipartite graph.

Post-processing estimation based on the conditional modes does not suffer from the fixed-effects estimator's dependence on the statistical properties of the observed graph. In particular, the influence matrix used to form the conditional modes is well-behaved, whereas the ill-conditioned influence matrix is the source of the post-processing bias affecting the fixed-effects estimator [Bates and Calderón, 2023; Kline, Saggio and Sølvsten, 2020b]. The mixed-effects estimator achieves these properties because it is a shrinkage estimator that pulls the conditional modes back to their assumed population mean thus limiting the influence of any particular match, which is also the objective of the KSS estimator for the fixed person and firm effects. Table 1 summarizes the methods that we have just discussed and indicates the exact reference for the estimators used in this paper.

| Table 1: Summary of Estimation Methods for AKM-Style Models | | | |
|---|---|---|---|
| Descriptor | Reference | Brief Description | In This Paper? |
| Original AKM | Abowd, Kramarz, Margolis [1999b] | Two-step approximate solution to fixed-effects estimator | No |
| Enhanced AKM | Abowd, Finer, Kramarz [1999a] | Improved-precision two-step approximate solution to fixed -effects estimator. | No |
| AKM | Abowd, Creecy, Kramarz [2002] | Exact solution to fixed-effects estimator | Yes |
| GLM | Bates et al. [2015] | Two step implementation of exact fixed-effects estimator for AKM using standard software | Yes |
| CorrME | Abowd, Kramarz [1999c]; Woodcock [2008] | Mixed-effects estimator with fixed covariate effects and random, person/firm effects where correlation is due to a match effect | No (see Appendix, Section D) |
| ME | Bates, Calderón, [2023]; Bates [2023] | High-performance mixed-effects estimator with fixed covariate effects and random, independent person/firm effects | Yes |
| KSS | Kline, Saggio, Sølvsten [2020] | Two-step solution to fixed-effects estimator with correction for estimated functions of the fixed person/firm effects | Yes |
| AMS | Abowd, McKinney, Schmutte [2019] | Dimension-reducing mixed-effects estimator based on network models | No |
| BLM | Bonhomme, Lamadon, Manresa [2022] | Dimension-reducing random effects estimator based on clustering | No |
| Notes: See Appendix, Section A for more details on the estimators. | | | |

This paper is organized as follows. Section 2 lays out the methods we use, as summarized in Table 1, with most details relegated to the Appendix, Section A. Section 3 is a detailed description of the data and the methods we use to construct the estimation samples. Section 4 presents our results. Section 5 discusses and offers recommendations.

1. Methods

The basic AKM specification contains time-invariant person, firm, and regression control effects shown for each individual and time period as

$$y_{it} = \theta_i + \psi_{J(i,t)} + x_{it}\beta + \varepsilon_{it}, \tag{1}$$

where the function $J(i,t)$ maps individual $i$ to their employer in period $t$; i.e., selects the correct employer from the bipartite graph of workers and firms. CHK and KSS also specify variants in which the employer effect has a time-varying structure as in

$$y_{it} = \theta_i + \psi_{J(i,t)t} + x_{it}\beta + \varepsilon_{it},$$

where the $t$ subscript on $\psi_{J(i,t)t}$ stands in for whatever function of time, usually nonoverlapping windows, is employed. Cahuc et al. [2006] derive a variant in which the equilibrium search generates a dependence on the firm effect from the previous employer, which can be represented as

$$y_{it} = \theta_i + \psi_{J(i,t)} + \psi_{J(i,t-\tau_{it})} + x_{it}\beta + \varepsilon_{it},$$

where the notation $\psi_{J(i,t-\tau_{it})}$ means select the employer effect from period $t - \tau_{it}$, the year in which individual $i$ separated from their most recent previous employer.

The statistical structure is the same for these models. We summarize it here for the basic AKM specification:

$i \in \{I^*\}$, where the set $I^*$ is a draw from the superpopulation of person effects,

$j \in \{J^*\}$, where the set $J^*$ is a draw from the superpopulation of firm effects,

$t = 1, \ldots, T$ arbitrary time window,

$\Omega = \{(I^*, J^*)\}$ is the superpopulation universe of person and firm effects.

In matrix notation the person- and firm-effects-only variant can be specified as

$$y = X\beta + D\theta + F\psi + \varepsilon,$$

where $y$, is the stacked vector of log real annual earnings, $X$ is the matrix of control variables conformably stacked, $D$ is the design of the unconstrained person effects, $F$ is the design of the unconstrained firm effects, $\beta$ are the fixed control variable effects, $\theta$ is the conformably stacked vector of person effects, $\psi$ is the conformably stacked vector of firm effects, and $\varepsilon$ is the conformably stacked vector of residuals. Estimable linear functions of the person and firm effects are formed by pre-multiplying the appropriate design matrix with the correct transformation, which is then applied to the estimated unconstrained effects in fixed-effects estimation. Any unconstrained fixed-effects solution can be used as the input to the estimable linear functions (identifiable parameters).

In the random effects specification, the stochastic assumptions are:

$$E\begin{bmatrix}\theta_i \\ \psi_j \\ \varepsilon_{it}\end{bmatrix} X, D, F = \begin{bmatrix}0 \\ 0 \\ 0\end{bmatrix} \forall (i,j) \in (I^*, J^*) \in \Omega, \forall t$$

$$V\begin{bmatrix}\theta_i \\ \psi_j \\ \varepsilon_{it}\end{bmatrix} X, D, F = \begin{bmatrix}\Sigma & 0 \\ 0 & \sigma_{\varepsilon\varepsilon}\end{bmatrix} \forall (i,j) \in (I^*, J^*) \in \Omega, \forall t$$

$$\Sigma = \begin{bmatrix}\sigma_{\theta\theta} & \sigma_{\theta\psi} \\ \sigma_{\theta\psi} & \sigma_{\psi\psi}\end{bmatrix} \text{ positive definite symmetric } \forall (I^*, J^*) \in \Omega$$

$$\sigma_{\varepsilon\varepsilon} > 0.$$

To form the likelihood function, used to form the profiled deviance, the objective function in mixed-effects estimation, we proceed sequentially. First, sample from the superpopulation

$(I^*, J^*) \leftarrow \Omega$ with probability drawn from $P(0, \Sigma)$.

to obtain a realization of potential workers and firms denoted $(I^*, J^*)$. Then, select from the population

$(I, J) \subset (I^*, J^*)$ active workers and firms.

Note that the key distinction between the draw from the superpopulation, which yields a network $(I^*, J^*)$ that is one realization of draws from $P(0, \Sigma)$, and the estimation population $(I, J)$ is that the resulting network is not a random sample from $(I^*, J^*)$; only the active jobs, and not the latent jobs, appear. In an experimental setting, a sample of the $(I^*, J^*)$ population could produce active jobs whose statistical properties match the superpopulation assumptions. However, in a non-experimental setting, $(I, J)$ is a selected subset of $(I^*, J^*)$ representing the job matches that actually occur. This means that even if $\sigma_{\theta\psi} = 0$ in the superpopulation distribution, the selected subpopulation could have any value of $\sigma_{\theta\psi}$.[3] In particular, it is meaningful to estimate the person and firm effects from the random-effects model because the realized graph $(I, J)$ contains only the employment relations (jobs) that the labor market realized, not a random subset from the superpopulation $(I^*, J^*)$. These can be used to estimate realized correlation in the sample but the superpopulation correlation, which applies to both observed and latent jobs remains unidentified. See Appendix, Section D. This is the primary distinction between the economic and biostatistical implementation of mixed-effects models.

The remainder of the derivation of the deviance function follows Bates [2023] exactly. His equation (6) is the exact likelihood, but it involves a high-dimensional integral. Estimation proceeds iteratively. Given an estimate of Σ, one computes the conditional modes (the estimated person and firm effects) and the conditional estimate of the regression coefficients. Then, using the profiled deviance (equation 21), one computes the maximum likelihood estimate of Σ. We used Julia [Bates and Calderón, 2023] implementations of lme4 to do the mixed-effects estimation; however, the computational details are similar to the R version [Bates et al. 2015]. We used our own implementation of ACK to do the fixed-effects estimation. See Appendix, Section A for further details.

## 2. Data

The empirical work in this paper uses earnings information from the Longitudinal Employer-Household Dynamics (LEHD) infrastructure files, developed and maintained by the U.S. Census Bureau [Abowd et al. 2009]. From this data source, we construct job-level annual earnings files using all available states for workers who are age 18-70, appear on the Census Numident, and never have an earnings year with more than 12 jobs.

In the LEHD data infrastructure, a job is the statutory employment of a worker by a statutory employer as defined by the Unemployment Insurance (UI) system in each state. Mandated reporting of UI-covered wage and salary payments between one statutory employer and one statutory employee is governed by the state's UI system. Reporting covers private employers and state and local government. There are no self-employment earnings unless the proprietor drew a salary, which, for UI earnings data, is indistinguishable from other employees.

Using our initial sample, we construct an estimate of total labor market experience using accumulated actual experience across all jobs anywhere in the full national LEHD data. We combine actual experience with an imputation of initial experience for workers who enter the LEHD data after age 18. We also impute annual hours worked at each job using a model developed from quarterly reported hours information available for a select number of states. We use the complete national LEHD infrastructure data to compute these regressors. See Abowd, McKinney and Zhou [2018] for details on their construction. Using the reported earnings information and the work history controls, we create Sample 1 by selecting the workers and firms active in Wyoming (WY), South Dakota (SD), and Montana (MT) during the years 1994-2017, inclusive. These three neighboring states were chosen due to their connected labor markets and their limited combined size, which is permits mixed-effects estimation without sampling, thus facilitating clean comparisons with fixed-effects estimators.

---
[3] The graph implied by this process does not resemble the one used in Abowd, McKinney and Schmutte [2019 because matches are reformed every period rather than evolving according to a dynamic process.

The estimation of mixed-effects models is computationally intensive; especially the memory required to store the design matrices $D$ and $F$. The usual approach in biostatistics, when estimates of the conditional modes are the objective, is to sample the input data, fit the mixed-effects model, then compute conditional modes for all data points in the full sample. This is an advantage of the mixed-effects methods when using large, linked employer-employee databases; however, we wanted to make direct comparisons using the same estimation samples throughout. So, we selected a small regional labor market within the U.S.

One way to reduce the memory required to fit the mixed effects models, while minimally affecting the results, is to discard persons and firms that contribute only a small amount of reliable information. We remove workers who are active in (WY, SD, MT) for less than one year or have more than 40 total jobs during the entire analysis period. We remove small firms—those with only one worker—and workers that primarily work in those small firms, defined as those that have up to 80% of their earnings in firms that employ at least two persons.[4] Finally, we retain only jobs in the largest connected component, which facilitates AKM estimation and the KSS correction but is not required for mixed-effects methods. These restrictions produced Sample 2. Table 2 shows the results of the sample restrictions.

| Table 2: Characteristics of the Analysis Samples | | | |
|---|---:|---:|---:|
| Sample | 1 | 2 | 3 |
| Observations | 40,120,000 | 38,230,000 | 5,088,000 |
| Jobs | 15,860,000 | 14,250,000 | 2,866,000 |
| Persons | 4,031,000 | 2,844,000 | 1,475,000 |
| Firms | 278,000 | 219,000 | 80,500 |
| Notes: Sample 1 consists of persons ages 18 to 70, inclusive, who are active in WY, SD, or MT during the years 1994-2017, inclusive, who appear on the Census Numident, and who have never had more than 12 jobs per year. The largest connected component of Sample 1 contains 99.9% of observations and jobs, 99.6% of persons, and 95.7% of firms. Sample 2 is a subset of the persons and firms in Sample 1 removing workers who are active for less than 1 year or who have more than 40 jobs. Sample 2 also removes firms with only 1 person and persons who primarily work in 1 firm. Finally, Sample 2 retains only person-firm pairs in the largest connected component. Sample 3 retains persons and firms from Sample 2 who are active during 2012-2014, inclusive and who are also in the 3-year largest connected component. ||||

Our final sample, Sample 3, is a subset of Sample 2 created by restricting the sample to workers active in 2012-2014, inclusive. We use the short 3-year Sample 3 to compare the estimation results across different model types with similar estimates constructed using our long 24-year Sample 2. The 3-year window was selected since KSS use such a window in their empirical work.

---

[4] Work in firms with only one employee is very rare. While this rule does exclude a worker who received 79% of earnings over the period from an employer with at least 2 workers, its main effect is to eliminate employers who had only one employee in the entire analysis period.

## 3. Results

Table 3 shows the basic estimation results. The column labeled AKM NR contains no regressors. The columns with the designation EH include a quartic for both experience and hours worked. The estimator used in the column GLM EH is a two-step fixed-effects estimator in which the fixed regression coefficients are estimated from the within-job specification (as in the original AKM paper) as step one. In step two, we use the residuals from step one to fit the fixed person and firm effects using ACK. The AKM EH column uses ACK to fit the fixed regressor, person and firm effects simultaneously. Finally, the MIX EH column uses the profiled deviance method to fit the fixed regressor coefficients. It uses the conditional modes to estimate the person and firm effects. Note that $\sigma_{\theta\psi} = 0$ in the mixed-effects estimator. See the Appendix, Section A for details.

**Table 3: Components of Log Real Annual Earnings Variance**

| Earnings Components | AKM NR | GLM EH | AKM EH | MIX EH |
|---|---|---|---|---|
| | Mean | | | |
| Log Real Annual Earnings ($y_{it}$) | 8.9110 | 8.9110 | 8.9110 | 8.9110 |
| Experience and Hours Effects ($x_{it}\beta$) | 0.0000 | 3.2260 | 3.5070 | 3.4750 |
| | Standard Deviation | | | |
| Log Real Annual Earnings ($y_{it}$) | 1.8490 | 1.8490 | 1.8490 | 1.8490 |
| Experience and Hours Effects ($x_{it}\beta$) | 0.0000 | 1.5740 | 1.6750 | 1.6840 |
| Composite Residual ($y_{it} - x_{it}\beta$) | 1.8490 | 0.5738 | 0.5419 | 0.5392 |
| Residual ($y_{it} - x_{it}\beta - \theta_i - \psi_{J(i,t)}$) | 1.2340 | 0.4149 | 0.4038 | 0.4101 |
| $\theta_i$ | 0.9570 | 0.2927 | 0.2729 | 0.2062 |
| $\psi_{J(i,t)}$ | 0.9287 | 0.2295 | 0.2014 | 0.2111 |
| Corr($\theta_i, \psi_{J(i,t)}$) | 0.0668 | 0.1398 | 0.1409 | 0.1725 |
| | Variance-Covariance Components | | | |
| Log Real Annual Earnings ($y_{it}$) | 3.4188 | 3.4188 | 3.4188 | 3.4188 |
| Experience and Hours ($x_{it}\beta$) | 0.0000 | 2.4775 | 2.8056 | 2.8359 |
| Cov($\theta_i, \psi_{J(i,t)}$) | 1.8970 | 0.1571 | 0.1305 | 0.1021 |
| Residual ($y_{it} - x_{it}\beta - \theta_i - \psi_{J(i,t)}$) | 1.5228 | 0.1721 | 0.1631 | 0.1682 |
| Other Covariance Terms | -0.0010 | 0.6121 | 0.3196 | 0.3127 |
| Total of Variance-Covariance Terms above | 3.4188 | 3.4188 | 3.4188 | 3.4188 |
| Share of Cov($\theta_i, \psi_{J(i,t)}$) in Total | 0.5549 | 0.0460 | 0.0382 | 0.0299 |

Notes: Estimates constructed using Sample 2. The dependent variable in all regression models is log real annual earnings (2015 CPI) at each job. The AKM NR model is estimated using no regression covariates, while the AKM EH model includes covariates. Models with covariates (EH) include a quartic in imputed-initial + actual experience and a quartic in imputed annual hours worked. The GLM EH model is a two-step AKM. In the first step, the covariate parameters are estimated using a within-job regression. In the second step, the residuals from the first step are used to estimate the AKM person ($\theta_i$) and firm ($\psi_{J(i,t)}$) components. The MIX EH model is estimated using a mixed-effects profiled deviance method with uncorrelated superpopulation person and firm variance components. The estimated person and firm variance components are 0.2410 and 0.2478, respectively. The MIX EH variance components are used to estimate conditional modes, the mixed-effects analogues of the AKM $\theta_i$ and $\psi_{J(i,t)}$. See Appendix, Section A for details.

The three models that contain regressors produce similar earnings components results in the mean, standard deviation, and variance panels. The main difference is the lower return to experience and hours worked (primarily experience) for the within-job GLM EH method. However, the larger positive correlation of the experience/hours effects with the person and firm effects, and the larger variance of those effects, offsets most of the effect of the lower return to experience. The correlations between $\theta_i$ and $\psi_{J(i,t)}$ are positive for all models, with the largest correlation shown for the MIX EH model. The long 24-year sample period likely plays a significant role in reducing the effects of limited mobility bias often seen in shorter samples.

In Table 4 we show the estimated person and firm effects variances as well as the covariance and correlation for different model types for both short (3-year) and long (24-year) samples. Using the 24-year sample, the MIX EH model has the lowest estimated plug-in person-effects variance, while the AKM EH model has the lowest estimated firm-effects variance. The MIX EH variance component estimates are noticeably larger, given that the conditional mode estimates of the person and firm effects are shrunk toward the mean of zero. The KSS-adjusted AKM person and firm effects variance estimates differ substantially between the leave out a match (KSS-match) and the leave out an observation (KSS-obs) approaches. The primary difference appears in the dramatically lower variance of the person-effects component. The extremely small person-effects variance results in an extremely large person-firm effects correlation estimate for the KSS-match approach compared with the KSS-obs approach (0.36 vs 0.18). If the KSS-match person effects variance were like the KSS-obs person-effects variance, then the person-firm effects correlation estimate would decrease to about 0.19. Somewhat surprisingly perhaps, the MIX EH plug-in estimate has the highest person-firm effects correlation among the non-KSS estimates. The information in the longer sample is extremely beneficial for the traditional approaches, reducing the error in the $\theta_i$ and $\psi_{J(i,t)}$ estimates, resulting in lower variance estimates and higher person-firm effects correlations relative to the shorter sample.

**Table 4: Variance and Covariance of Earnings Components Using Alternative Estimation Methods**

| | Sample 2: 24-Year Sample (1994-2017) | | | | |
|---|---|---|---|---|---|
| | Variance | | Person-Firm | | |
| Model | Person | Firm | Cov | Corr | Share |
| AKM NR Plug-in | 0.9158 | 0.8625 | 0.0593 | 0.0668 | 0.5549 |
| AKM EH Plug-in | 0.0745 | 0.0406 | 0.0077 | 0.1409 | 0.0382 |
| GLM EH Plug-in | 0.0857 | 0.0527 | 0.0094 | 0.1398 | 0.0460 |
| MIX EH Plug-in | 0.0425 | 0.0446 | 0.0075 | 0.1725 | 0.0299 |
| MIX EH Variance Components | 0.0581 | 0.0614 | 0.0000 | 0.0000 | 0.0349 |
| | | | | | |
| AKM NR KSS Leave Out Match | 0.6166 | 0.7809 | 0.0825 | 0.1188 | 0.4570 |
| AKM NR KSS Leave Out Observation | 0.7955 | 0.8454 | 0.0673 | 0.0821 | 0.5193 |
| | | | | | |
| AKM EH KSS Leave Out Match | 0.0171 | 0.0381 | 0.0092 | 0.3607 | 0.0215 |
| AKM EH KSS Leave Out Observation | 0.0598 | 0.0386 | 0.0085 | 0.1776 | 0.0338 |
| | Sample 3: 3-Year Sample (2012-2014) | | | | |
| | Variance | | Person-Firm | | |
| Model | Person | Firm | Cov | Corr | Share |
| AKM NR Plug-in | 1.4982 | 0.9928 | -0.0445 | -0.0365 | 0.7219 |
| AKM EH Plug-in | 0.1442 | 0.0487 | 0.0035 | 0.0413 | 0.0601 |
| GLM EH Plug-in | 0.1628 | 0.0633 | 0.0075 | 0.0735 | 0.0724 |
| MIX EH Plug-in | 0.0375 | 0.0612 | 0.0058 | 0.1208 | 0.0331 |
| MIX EH Variance Components | 0.0674 | 0.0707 | 0.0000 | 0.0000 | 0.0415 |
| | | | | | |
| AKM NR KSS Leave Out Match | 0.5134 | 0.7860 | 0.1072 | 0.1687 | 0.4550 |
| AKM NR KSS Leave Out Observation | 1.0750 | 0.8828 | 0.0431 | 0.0442 | 0.6144 |
| | | | | | |
| AKM EH KSS Leave Out Match | 0.0858 | 0.0369 | 0.0114 | 0.2028 | 0.0437 |
| AKM EH KSS Leave Out Observation | 0.0956 | 0.0376 | 0.0108 | 0.1807 | 0.0465 |

Notes: The dependent variable is log real annual earnings (2015 CPI) at each job. The AKM NR model is estimated using no regression covariates, while the AKM EH model includes covariates. Models with covariates (EH) include a quartic in imputed-initial + actual experience and a quartic in imputed annual hours worked. The GLM EH model is a two-step AKM EH implementation. In the first stage, the covariate parameters are estimated using a within-job regression. In the second stage, the residuals from the first stage are used to estimate the AKM person ($\theta_i$) and firm ($\psi_{J(i,t)}$) components. The MIX EH model is estimated using a mixed-effects profiled deviance method with uncorrelated superpopulation person and firm variance components. The MIX EH variance component estimates are used to compute the conditional modes, the mixed-effect analogue of the AKM $\theta_i$ and $\psi_{J(i,t)}$, which are then used in the MIX EH plug-in rows. The plug-in variance and covariance estimates are calculated directly from the estimated person ($\theta_i$) and firm ($\psi_{J(i,t)}$) effects. The KSS estimates are adjusted using the methods of Kline, Saggio, and Sølvsten [2020]. The Corr column shows the correlation between the person and firm earnings components. The Share column shows the sum of the person and firm earnings variance plus two times the covariance all divided by the earnings variance. See Appendix, Section A for estimation details and Section C for standard error information.

In Table 5 we try to resolve the discrepancies across the various approaches by creating simulated data. We vary the correlation between the person and firm effects as well as the number of observations per match (job). In the simulated data there are 2,000 workers, each of whom works for $T$ periods, which is a relatively small sample compared to those used in the literature. There are 40 firms. We first draw a person-effect earnings component from a $N(0, \sigma_p)$. Next, we draw a firm-effect earnings component from a $N(0, \sigma_f)$, conditional on the person earnings component and the specified correlation. We draw two firm components for each worker, sampling with replacement. Next, we divide the support of the realized firm-effects component distribution into 40 equally likely non-overlapping regions. These regions represent the 40 different types of firms. Each person works $T$ periods, $T/2$ in the first firm and $T/2$ in the second firm. Depending on the two firm-effects draws, the person is either in the same firm in both periods if the draws are close or in another firm if they differ by more than the firm bin width (the bin width varies depending on its location in the distribution to keep the firm sizes roughly equal). The higher the person-firm effects correlation, the more likely it is that a person works in two high (low) earning firms. The dependent variable, log earnings is equal to $10 + earn_p + earn_f + earnexp + earnresid$. The standard deviation of person effects is 0.26, standard deviation of firm effects is 0.25, and the standard deviation of the residual is 0.39. Initial experience is chosen from a Poisson distribution with a $l$=4 and support truncated at a maximum of 24.

Although well suited for comparisons across methods, our simulated data may not be the best choice to disentangle all the differences between KSS-match and KSS-obs. There is no serial correlation in the residuals and the graph is well-connected with equal duration jobs for all workers. However, there are some workers that only have one employer which affects both the KSS-match and KSS-obs estimates. The main intent of the simulation was to study how the correlation between person and firm effects and the number of time periods in the sample are related when the bipartite graph is well-behaved. In the simulations we produce plug-in estimates using fixed-effects estimates via the AKM model and mixed-effects estimates via the profiled deviance method. We also produce the KSS-match and KSS-obs estimates. All models include a quartic in experience.

Table 5 shows that for moderate levels of correlation between the person and firm effects, the biases in the plug-in estimates in the simulation are smaller than one might expect from the KSS critique. Results improve as the number of time periods increases, as Bonhomme et al. also find. The fixed-effects plug-in estimator overestimates the person-effects variance. The mixed-effects estimates are a bit more interesting. The plug-in person-effect variances are underestimated even when there is no correlation with the firm effect. Person-firm correlation has a larger impact on the mixed-effects estimates relative to their fixed-effects counterparts, especially in short samples. For the longer samples, the plug-in results are reasonable when the true correlation is less than about 0.4. The KSS-match method slightly overestimates the person variance but both the KSS-match and KSS-obs methods are otherwise stable for the relatively small samples used in the simulation regardless of the true correlation. The simulation results are fully consistent with the published KSS results and support the conclusion that their method can reliably estimate the covariance matrix Σ even in samples using only a few years of data.

What the simulation does not address, however, is the reliability of the estimated person and firm effects when using long versus short samples. Table 5 demonstrates that conclusions about the person effect are strongly affected by using short samples. These short samples are often part of research designs like those in CHK where variation in the firm effects over time is a potentially important possibility. Our results suggest that using long samples but allowing the firm effect to time-vary should give more reliable results for the person effect because it does not limit the information used by creating nonoverlapping subsamples. Mixed-effects estimation is easier to implement this way because it does not require use of the largest connected component in the bipartite graph. The plug-in estimates from the mixed effects model are also not subject to the estimation bias that KSS demonstrate for the fixed-effects method when estimating $\Sigma$. Instead, any bias in the mixed-effects estimates arises from shrinking the estimated conditional modes back to the grand mean of zero.

| Table 5: Standard Deviations and Correlations of Person and Firm Effects in Simulated Data | | | | | | | | | |
|---|---|---|---|---|---|---|---|---|---|
| Fixed Effects Using AKM | | | | | | | | | |
| Sim | 6 periods | | | 12 periods | | | 24 periods | | |
| Corr | Person | Firm | Corr | Person | Firm | Corr | Person | Firm | Corr |
| 0.0 | 0.310 | 0.250 | -0.006 | 0.287 | 0.250 | -0.008 | 0.277 | 0.250 | 0.002 |
| 0.2 | 0.310 | 0.250 | 0.165 | 0.287 | 0.251 | 0.183 | 0.277 | 0.249 | 0.186 |
| 0.4 | 0.312 | 0.250 | 0.333 | 0.288 | 0.250 | 0.359 | 0.278 | 0.247 | 0.374 |
| 0.6 | 0.312 | 0.249 | 0.500 | 0.289 | 0.246 | 0.544 | 0.278 | 0.247 | 0.567 |
| 0.8 | 0.312 | 0.248 | 0.674 | 0.291 | 0.247 | 0.731 | 0.278 | 0.249 | 0.757 |
| Mixed Effects | | | | | | | | | |
| Sim | 6 periods | | | 12 periods | | | 24 periods | | |
| Corr | Person | Firm | Corr | Person | Firm | Corr | Person | Firm | Corr |
| 0.0 | 0.227 | 0.249 | 0.002 | 0.242 | 0.247 | 0.001 | 0.254 | 0.249 | 0.000 |
| 0.2 | 0.223 | 0.273 | 0.135 | 0.239 | 0.263 | 0.158 | 0.253 | 0.257 | 0.173 |
| 0.4 | 0.208 | 0.299 | 0.264 | 0.230 | 0.282 | 0.321 | 0.246 | 0.268 | 0.353 |
| 0.6 | 0.169 | 0.348 | 0.343 | 0.203 | 0.318 | 0.443 | 0.232 | 0.287 | 0.515 |
| 0.8 | 0.092 | 0.434 | 0.284 | 0.124 | 0.408 | 0.423 | 0.177 | 0.349 | 0.616 |
| Fixed Effects with the KSS (match) Correction | | | | | | | | | |
| Sim | 6 periods | | | 12 periods | | | 24 periods | | |
| Corr | Person | Firm | Corr | Person | Firm | Corr | Person | Firm | Corr |
| 0.0 | 0.265 | 0.250 | 0.006 | 0.263 | 0.248 | 0.000 | 0.265 | 0.248 | 0.000 |
| 0.2 | 0.266 | 0.248 | 0.217 | 0.263 | 0.250 | 0.203 | 0.266 | 0.249 | 0.196 |
| 0.4 | 0.266 | 0.249 | 0.400 | 0.264 | 0.248 | 0.403 | 0.266 | 0.247 | 0.395 |
| 0.6 | 0.268 | 0.245 | 0.602 | 0.264 | 0.248 | 0.597 | 0.267 | 0.247 | 0.593 |
| 0.8 | 0.269 | 0.242 | 0.800 | 0.268 | 0.243 | 0.799 | 0.268 | 0.246 | 0.791 |
| Fixed effects with the KSS (observation) Correction | | | | | | | | | |
| Sim | 6 periods | | | 12 periods | | | 24 periods | | |
| Corr | Person | Firm | Corr | Person | Firm | Corr | Person | Firm | Corr |
| 0.0 | 0.260 | 0.252 | 0.002 | 0.260 | 0.249 | -0.003 | 0.260 | 0.249 | -0.007 |
| 0.2 | 0.260 | 0.249 | 0.191 | 0.260 | 0.250 | 0.206 | 0.260 | 0.248 | 0.199 |
| 0.4 | 0.260 | 0.250 | 0.399 | 0.260 | 0.250 | 0.400 | 0.260 | 0.250 | 0.401 |
| 0.6 | 0.260 | 0.250 | 0.601 | 0.260 | 0.250 | 0.602 | 0.260 | 0.250 | 0.601 |
| 0.8 | 0.260 | 0.249 | 0.800 | 0.260 | 0.251 | 0.800 | 0.260 | 0.250 | 0.801 |

Notes: All earnings models are estimated using simulated data with 2,000 persons and 40 firms. Each person has one job per year and the total number of observations are split equally between either one or two potential employers (person-firm matches are made with replacement). A quartic in experience is included in all models. The simulated data has a person-effect standard deviation (SD) of 0.26 (variance = 0.0676) and a firm-effect SD of 0.25 (variance = 0.0625) with varying levels of correlation (Sim Corr) between the person and firm effects. The residuals have no induced serial correlation or heteroscedasticity. Each combination of model type (fixed effects with AKM, mixed effects, and fixed effects with KSS (match and observation correction), number of time periods (6, 12, 24), and correlation (0.0, … ,0.8) shows the median person-effect SD, firm-effect SD, and person-firm correlation across the 25 simulations. Fixed effects with AKM and mixed effects panels are plug-in estimates.

## 5. Discussion and Recommendations

We use long and short samples of workers and firms from three western states in the U.S. to compare fixed- and mixed-effects estimators of the AKM person and firm effects. We compare plug-in estimates of the covariance matrix of the person and firm effects to those calculated using the KSS correction. The KSS correction reliably estimates that covariance matrix. We note, however, that the mixed-effects plug-in estimates do not suffer from the bias noted in KSS. Any bias in the mixed-effects plug-in estimators is a consequence of the shrinkage that the profiled deviance method imposes when computing the conditional modes, which substitute for the fixed person and firm effects in plug-in estimators.

We recommend considering the use of mixed-effects estimators when the correlation between the person and firm effects is less than 0.4, a criterion which is met by every non-simulated estimate in this paper. The mixed-effects estimator produces reliable estimates of the person effect variance for both short and long samples, while also producing reasonable estimates of the firm variance and the person-firm correlation. We also suggest using both fixed- and mixed-effects methods to model temporal variation in the firm effect and to avoid using short temporal samples to estimate person effects.

These recommendations are meant to be practical. If a researcher is working with a universe sample of jobs, it is often the case that all estimators require more computational resources than are available, particularly if the research is being conducted in a supervised-use environment controlled by a national statistical office. Sampling by person or firm and using the mixed-effects estimator should be more reliable than using any of the fixed-effects estimators, since the bipartite graph of the sampled data will be very disconnected. The KSS-match estimator must then operate on a selected subset of the sample and may give very misleading results, although the KSS-obs estimator can still be used along with the ACK or other iterative fixed-effects solutions. More importantly, if the mixed-effects estimator is used, the software can produce estimated person and firm effects (conditional modes) for the universe without re-estimating the entire model.

## References

ABOWD, J, FINER, H. and KRAMARZ, F. [1999a] "Individual and Firm Heterogeneity in Compensation: An Analysis of Matched Longitudinal Employer-Employee Data for the State of Washington" in J. Haltiwanger et al. (eds.) *The Creation and Analysis of Employer-Employee Matched Data*, (Amsterdam: North Holland), pp. 3-24.

ABOWD, J., KRAMARZ, F. and MARGOLIS, D. [1999b] "High Wage Workers and High Wage Firms," *Econometrica*, Vol. 67(2):251-333. DOI: https://doi.org/10.1111/1468-0262.00020.

ABOWD, J. and KRAMARZ, F. [1999c] "Econometric Analysis of Linked Employer-Employee Data," *Labour Economics*, Vol. 6 (March):53-74. DOI: https://doi.org/10.1016/S0927-5371(99)00003-2.

ABOWD, J., CREECY, R. and KRAMARZ, F. [2002] "Computing Person and Firm Effects Using Linked Longitudinal Employer-Employee Data," U.S Census Bureau, Technical Paper TP-2002-06. URL: https://www2.census.gov/ces/tp/tp-2002-06.pdf (retrieved May 16, 2023).

ABOWD, J., STEPHENS, B., VILHUBER, L., ANDERSSON, F., MCKINNEY, K., ROEMER, M. and WOODCOCK, S. [2009] "The LEHD infrastructure files and the creation of the Quarterly Workforce Indicators." In *Producer dynamics: New evidence from micro data*, Dunne, T., Jensen, J.B. and Roberts, M. (eds.), pp. 149–230. Chicago: University of Chicago Press. http://www.nber.org/chapters/c0485.

ABOWD, J., MCKINNEY, K. and ZHAO, N. [2018] "Earnings Inequality and Mobility Trends in the United States: Nationally Representative Estimates from Longitudinally Linked Employer-Employee Data," *Journal of Labor Economics* 36, S1 (January):S183-S300. DOI: https://doi.org/10.1086/694104

ABOWD, J., MCKINNEY, K. and Schmutte, I. [2019] "Modeling Endogenous Mobility in Earnings Determination," *Journal of Business and Economic Statistics*, Vol. 37(3):405-418. DOI: https://doi.org/10.1080/07350015.2017.1356727.

BATES, D., MÄCHLER, M., BOLKER, B. and WALKER, S. [2015] "Fitting Linear Mixed-Effects Models Using lme4*," Journal of Statistical Software*, Vol. 67(1):1–48. DOI: https://doi.org/10.18637/jss.v067.i01 Software: https://cran.r-project.org/web/packages/lme4/ (retrieved on May 17, 2023).

BATES , D. [2023] "Computational methods for mixed models," CRAN library. URL: https://cran.r-project.org/web/packages/lme4/vignettes/Theory.pdf. (retrieved May 17, 2023)

BATES , D. and CALDERÓN, J.B.S. [2023] A Julia package for fitting (statistical) mixed-effects models. Software: https://github.com/JuliaStats/MixedModels.jl. (retrieved May 17, 2023)

BONHOMME, S., LAMADON, T. and MANRESA, E. [2022] "Discretizing Unobserved Heterogeneity," *Econometrica*, Vol. 90(2):625-643. DOI: https://doi.org/10.3982/ECTA15238

BONHOMME, S.,, HOLZHEU, K., LAMADON, T., MANRESA, E., MOGSTAD, M. and SETZLER B. [2022] "How Much Should We Trust Estimates of Firm Effects and Worker Sorting?" *Journal of Labor Economics*, Vol. 41(2):291-322.DOI: https://doi.org/10.1086/720009

CAHUC, P., POSTEL-VINAY, F. and ROBIN, J-M. [2006] "Wage Bargaining with On-the-job Search: Theory and Evidence," *Econometrica*, Vol. 74:323-364. DOI: https://doi.org/10.1111/j.1468-0262.2006.00665.x

HENDERSON, C. [1950] "Estimation of genetic parameters (abstract)," *Annals of Mathematical Statistics*, Vol. 21:309–310. DOI: https://doi.org/10.1214/aoms/1177729851

HENDERSON , C. [1953] "Estimation of variance and covariance components," *Biometrics*, Vol. 9:226–252. DOI: https://doi.org/10.2307/3001853

KLINE, P., SAGGIO, R. and SØLVSTEN, M. [2020a] "Leave-out Estimation of Variance Components," *Econometrica*, Vol. 88(5):1859-1898. DOI: https://doi.org/10.3982/ECTA16410

KLINE, P., SAGGIO, R. and SØLVSTEN, M. [2020b] "Variance Components HDFE" Julia package. URL: https://github.com/HighDimensionalEconLab/VarianceComponentsHDFE.jl; vignette (MatLab version only): https://github.com/rsaggio87/LeaveOutTwoWay/blob/master/doc/VIGNETTE.pdf. (retrieved July 31, 2023)

MORTENSEN, D. [2005] *Wage Dispersion: Why Are Similar Workers Paid Differently?* MIT Press. ISBN 9780262633192

POSTEL-VINAY, F. and ROBIN, J-M. [2002] "Equilibrium Wage Dispersion with Worker and Employer Heterogeneity," *Econometrica*, Vol. 70(6):2295-2350. DOI: https://doi.org/10.1111/j.1468-0262.2002.00441.x

SEARLE, S. [1971] *Linear Models*, Wiley ISBN-13 978-0471184997. Wiley-Interscience version (1997). DOI: https://doi.org/10.1002/9781118491782

WOODCOCK, S. [2008] "Wage Differentials in the Presence of Unobserved Worker, Firm, and Match Heterogeneity," *Labour Economics*, Vol. 15(4):771-793. DOI: https://doi.org/10.1016/j.labeco.2007.06.003

Appendix

## A. Computational formulas for estimators used in tables

AKM-NR: The model specification is equation (1). The regressors include only a constant. Run the ACK algorithm to produce any solution. Post-process the estimated person effects to have zero mean by observation. Post-process the estimated firm effects to have zero mean by observation.

AKM-EH: The model specification is equation (1). The regressors include the experience and hours variables described in the main text. Run the ACK algorithm to produce any solution. Post-process the estimated person effects to have zero mean by observation. Post-process the estimated firm effects to have zero mean by observation.

GLM EH: Transform equation (1) by computing deviations from within job means:

$$\bar{y}_{ij} = \frac{\sum_{t=1}^{T} \sum_{J(i,t)=j} y_{it}}{\sum_{t=1}^{T} \sum_{J(i,t)=j} 1(J(i,t)=j)}$$

$$\bar{x}_{ij} = \frac{\sum_{t=1}^{T} \sum_{J(i,t)=j} x_{it}}{\sum_{t=1}^{T} \sum_{J(i,t)=j} 1(J(i,t)=j)}$$

$$y_{it} - \bar{y}_{ij} = (x_{it} - \bar{x}_{ij})\beta + \varepsilon_{it}. \tag{2}$$

Estimate equation (2) using the same regressors as in AKM-EH by ordinary least squares to obtain $\hat{\beta}_{\text{GLM}}$. Form the composite residual as $e_{it} = y_{it} - x_{it}\hat{\beta}_{\text{GLM}}$. Run the ACK algorithm on the composite residual to obtain any solution. Post-process the estimated person effects to have zero mean by observation. Post-process the estimated firm effects to have zero mean by observation.

MIX-EH: In the Julia package [Bates and Calderón, 2023], specify the regressors as fixed effects using the same list as in AKM-EH, specify the person, firm and residual as independent normal random effects with zero means and unknown variances. Use the maximum likelihood estimator to estimate the coefficients on the regressors ($\beta$) and the variances of the random effects. Output the estimated conditional modes for the person and firm effects. In principle, no post-processing is required because the conditional modes have zero mean over persons and firms, respectively; however, to make the MIX-EH estimates strictly comparable to the AKM and KSS estimates, we post-processed the person and firm effects to have zero mean by observation.

KSS-obs: The model specification is equation (1). Compute the composite residual from the AKM-EH estimate as $e_{it} = y_{it} - x_{it}\hat{\beta}_{\text{AKM-EH}}$. Using the Julia implementation of KSS [Kline et al. 2020b], input the composite residual, specify the by-observation leave-one-out estimator, accept defaults elsewhere. Statistics that appear in tables are output from the procedure.

KSS-match: Using the same setup as in KSS-obs, specify the by-match leave-one-out estimator, accept the defaults elsewhere. Statistics that appear in tables are output from the procedure.

## B. The mixed-effects estimator does not suffer from limited mobility bias

Limited-mobility bias arises from properties of the bipartite graph connecting the jobs in the estimation sample of the ACK (exact fixed-effects) estimator. The adjacency matrix of this graph is block diagonal. Each block is called a connected group of jobs. The block containing the largest number of jobs is called the largest connected group of persons and firms. Within the blocks of the adjacency matrix, identification of the fixed person and firm effects (estimable functions in statistics) requires reducing the dimensionality of each effect by at least one—standard conditions for estimating the coefficients of indicator variables. To estimate across the blocks, that is, to use all the jobs in the estimation, requires at least one additional dimension reduction allocating the effect captured by the between-connected-groups variation. For this reason, and because in modern samples the largest connected group often contains the overwhelming majority of jobs, most applied work uses only the largest connected group (as we do in this paper).

Within the largest connected group, limited-mobility bias arises from jobs that are only connected to the rest of the group by a single edge—one person or firm in common with the rest of the persons and firms in the group because of that edge. Dropping that edge would create a new connected group disjoint from the largest one. The presence of these edges in the bipartite graph creates estimation instability in the fixed-effects algorithms because the observations associated with them cause singularities in the influence matrix. The KSS leave-out-one estimators correct for this instability. However, the KSS-match estimator can only be computed using the persons who have more than one employer in the estimation sample. This is only a subset, and sometimes a very small highly selected one for short estimation periods, of the observations for which the fixed person and firm effects are identified (estimable in statistics).

The mixed-effects estimator works with the entire graph, although we limited the estimation in this paper to the largest connected group for comparability to the other methods. Neither identification nor computation of the estimated conditional modes (the mixed-effects estimator comparable to the fixed-effects person and firm estimator) relies on the properties of the influence matrix. The influence matrix is always bounded. For mixed-effects methods, there are no convenient computational formulas for either the influence matrix or leave-one-out estimators. The random effects estimated by the mixed-effects algorithms (conditional modes) are the maximum likelihood estimates when the distribution of all random effects is normal. Thus, (bounded, continuous, differentiable) functions of these effects are also maximum likelihood estimates. There is not an incidental parameter problem because the conditional modes depend only on the finite set of regression coefficients and covariance parameters, and on the asymptotic features of the linked employee-employer data. (See Bates and Calderón [2023] and Bates [2023] for theoretical and computational details.)

The price of using the mixed-effects estimator instead of the fixed-effects estimator is the bias induced by shrinking all person and firm effects to the grand mean, usually set to zero to mimic the assumptions used in fixed-effect estimators. There is no closed-form solution to the Henderson [1953] equations. If there were, there would also be a closed-form solution to influence matrix and, therefore, a closed form solution to the estimated variance of each person and firm conditional mode.

## C.    *Standard errors for estimated functions of the person and firm effects*

In this paper, as in CHK and KSS, we use the universe of observations in all computations. Finite population standard errors, conventionally used by statistical agencies when estimating directly from a universe sample, are essentially zero for all model effects as in Abowd, McKinney and Zhao [2018] Table E7, which shows regression coefficients from an AKM specification very similar to the one used in this paper. However, such finite-population standard errors do not account for edit, estimation, confidentiality protection and modeling uncertainty. McKinney et al. [2021] thoroughly investigated the total variability of estimates based on the LEHD infrastructure edit, imputation, confidentiality protections and modeling sources. For cells as large as the ones used in this paper, such variability is unlikely to affect either the third or fourth significant digit, and Census Bureau publication rules limit our tables to four significant digits. For all practical purposes, reported computed statistics and differences between them are "statistically significant" at any conventional level; however, small differences in magnitude are not economically meaningful.

## D. *The superpopulation correlation between the person and firm effects is not estimable*

We demonstrate here that when estimating the correlation between the person and firm effects using the profiled deviance mixed-effects method, the matrix $\Sigma$ in the main text, it is necessary to also force zero correlation between the firm effects. In general, this will limit the estimable range of correlations between the person and firm effects.

In the random effects part of the design, denote the variance-covariance matrix between all person and firm effects as:

$$VCM \equiv C = \begin{bmatrix} A & B^T \\ B & D \end{bmatrix},$$

where

$$A = \sigma_p^2 I_{n_p}, B = \sigma_{pf} 1_{n_f,n_p}, D = \sigma_f^2 I_{n_f}, \rho = \frac{\sigma_{pf}}{\sigma_p \sigma_f},$$

$n_p$ = number of persons, $n_f$ = number of firms.

Then, factor as

$$C = \begin{bmatrix} I & 0 \\ BA^{-1} & I \end{bmatrix} \begin{bmatrix} A & 0 \\ 0 & S \end{bmatrix} \begin{bmatrix} I & A^{-1}B^T \\ 0 & I \end{bmatrix},$$

where

$$S = D - BA^{-1}B^T$$

$$= \begin{bmatrix} L_A & 0 \\ B(L_A^{-1})^T & L_S \end{bmatrix} \begin{bmatrix} L_A^T & L_A^{-1}B^T \\ 0 & L_S^T \end{bmatrix} = L_C L_C^T.$$

Note that

$$L_A = \sigma_p I_{n_p}, \ L_A L_A^T = \sigma_p^2 I_{n_p}, \ L_A^{-1} = (L_A^{-1})^T = \sigma_p^{-1} I_{n_p}$$

and

$$S = \sigma_f^2 I_{n_f} - \frac{\sigma_{pf}^2}{\sigma_p^2} n_p 1_{n_f,n_f} = \sigma_f^2 \left[ I - \rho^2 n_p 1_{n_f,n_f} \right].$$

$L_S$ can be solved recursively [Chen et al. 2013]. The solution for $(n_p, 2)$ is:

$$L_S = \sigma_f \begin{bmatrix} \sqrt{1 - \rho^2 n_p} & 0 \\ \dfrac{-\rho^2 n_p}{\sqrt{1 - \rho^2 n_p}} & \sqrt{1 - \rho^2 n_p + \dfrac{\rho^4 n_p}{1 - \rho^2 n_p}} \end{bmatrix}$$

$$L_S^T = \sigma_f \begin{bmatrix} \sqrt{1-\rho^2 n_p} & \dfrac{-\rho^2 n_p}{\sqrt{1-\rho^2 n_p}} \\ 0 & \sqrt{1-\rho^2 n_p + \dfrac{\rho^4 n_p^2}{1-\rho^2 n_p}} \end{bmatrix}.$$

Since

$$\left(\dfrac{-\rho^2 n_p}{\sqrt{1-\rho^2 n_p}}\right)^2 + 1 - \rho^2 n_p + \dfrac{\rho^4 n_p^2}{1-\rho^2 n_p} = 1 - \rho^2 n_p,$$

we have

$$L_S L_S^T = \sigma_f \begin{bmatrix} 1-\rho^2 n_p & -\rho^2 n_p \\ -\rho^2 n_p & 1-\rho^2 n_p \end{bmatrix} = S$$

and

$$L_C L_C^T = \begin{bmatrix} L_A L_A^T & L_A L_A^{-1} B^T \\ B(L_A^{-1})^T L_A^T & B(L_A^{-1})^T L_A^{-1} B^T + L_S L_S^T \end{bmatrix}.$$

Since

$$B(L_A^{-1})^T L_A^{-1} B^T + L_S L_S^T = BA^{-1} B^T + D - BA^{-1} B^T = D$$

and

$$L_C L_C^T = \begin{bmatrix} \sigma_p^2 I_{n_p} & \left[\sigma_{pf} 1_{n_f, n_p}\right]^T \\ \sigma_{pf} 1_{n_f, n_p} & \sigma_f^2 I_{n_f} \end{bmatrix} = C.$$

Therefore, in the profiled deviance, which is essentially the empirical expected $-2log(\mathcal{L})$ under the assumptions of the model (independence of the random effects from the design of the fixed effects), there is no way to estimate $\rho$. As $, n_p \to \infty$, the correlation $\rho$ between the person and firm effects must go to 0 to keep the matrix $C$ positive semi-definite. Nonzero correlation results from the realized design of the person and firm effects in the estimation sample; that is, from the absence of the latent jobs that are part of the superpopulation, and can only be estimated empirically from the realized design. We verified this using the same simulation design as in the main text.

## Additional references

MCKINNEY, K., GREEN, A., VILHUBER, L. and ABOWD, J. [2021] "Total Error and Variability Measures for the Quarterly Workforce Indicators and LEHD Origin-Destination Employment Statistics in OnTheMap" *Journal of Survey Statistics and Methodology* , Vol. 9(5): 1146–1182. DOI: https://doi.org/10.1093/jssam/smaa029, Supplemental online materials DOI: https://doi.org/10.5281/zenodo.3951670

CHEN, J., JIN, Z., SHI, Q., QIU, J. and LIU W. [2013]. "Block Algorithm and Its Implementation for Cholesky Factorization" *ICCGI 2013: The Eighth International Multi-Conference on Computing* *in the Global Information Technology*, pp. 232-6. ISBN: 978-1-61208-283-7 URL: http://personales.upv.es/thinkmind/dl/conferences/iccgi/iccgi_2013/iccgi_2013_11_30_10133.pdf (retrieved June 19, 2023)